\documentclass{Interspeech}

 \interspeechcameraready %

\title{Direction-Aware Neural Acoustic Fields for Few-Shot Interpolation\\of Ambisonic Impulse Responses}

\author[affiliation={1,2}]{Christopher}{Ick}
\author[affiliation={1}]{Gordon}{Wichern}
\author[affiliation={1}]{Yoshiki}{Masuyama}
\author[affiliation={1}]{François}{Germain}
\author[affiliation={1}]{Jonathan}{Le Roux}

\affiliation{}{Mitsubishi Electric Research Laboratories (MERL)}{Cambridge, MA, USA}
\affiliation{Music and Audio Research Laboratory}{New York University}{Brooklyn, NY, USA}
\email{chris.ick@nyu.edu \quad \{wichern,masuyama,germain,leroux\}@merl.com}

\usepackage{comment}
\usepackage{cite}

\begin{document}

\maketitle

\begin{abstract}
    The characteristics of a sound field are intrinsically linked to the geometric and spatial properties of the environment surrounding a sound source and a listener.
    The physics of sound propagation is captured in a time-domain signal known as a room impulse response (RIR).
    Prior work using neural fields (NFs) has allowed learning spatially-continuous representations of RIRs from finite RIR measurements.
    However, previous NF-based methods have focused on monaural omnidirectional or at most binaural listeners, which does not precisely capture the directional characteristics of a real sound field at a single point.
    We propose a direction-aware neural field (DANF) that more explicitly incorporates the directional information by Ambisonic-format RIRs.
    While DANF inherently captures spatial relations between sources and listeners, we further propose a direction-aware loss. In addition, we investigate the ability of DANF to adapt to new rooms in various ways including low-rank adaptation.
\end{abstract}

\section{Introduction}
Inferring how an acoustic signal emitted from a sound source is received by a listener involves modeling the physically complex propagation of that signal through their shared environment~\cite{roginska18immersive}.
Even when the source and listener are in direct line-of-sight, reflections of the sound signal at various surfaces and objects add significant contributions to what the listener hears.
Precisely capturing the acoustic properties of this process, which is typically done in the form of a room impulse response (RIR) to be convolved with a dry signal at the source, is crucial for accurate sound rendering, which in turn enables immersive experiences in media, particularly in virtual and augmented reality~\cite{Zaunschirm2018, Rana2019}.

Because the process of recording RIRs can be time-consuming and impractical in certain settings, there exists a broad set of methods for simulating RIRs, typically either by 
geometric acoustics~\cite{ISM, Scheibler2018PyRoom, chen20soundspaces},
or by wave-based methods~\cite{thompson2006wavereview, mehra2014waveprop, mehra2021wave, ChenJeng-Tzong2010Aomr}.
Each method has different benefits depending on the application, but both require an accurate geometric and material specification of a given scene to produce accurate RIRs.

The problem of interpolating RIRs from a finite number of measurements has been widely studied in a variety of settings, including
wave-based models \cite{raghuvanshi2014parametricwave,Remi2014} or geometric models \cite{raghuvanshi2018parametricgeo,Tsunokuni2021}.
In the particular case of learning spatially continuous RIRs for a scene with varying source/listener positions, inspiration has been taken from novel-view synthesis from the vision community \cite{mildenhall2020nerf} to develop continuous representations of acoustic scenes using neural fields (NFs) \cite{luo22naf, su22inras, liang23avnerf, liang2023nacf, chen2024RAF} and physics-informed neural networks~\cite{pezzoli2023implicit,Karakonstantis2024,koyama2024pinnsound}.
These methods draw from prior work in omnidirectional single-channel IR estimation, in which room acoustic metrics (such as T60, C50, and EDT) play a large role in perceptual accuracy, and have mainly evaluated binaural RIRs by averaging metrics across each channel.
However, these approaches fail to account for inter-channel directional information, which are crucial not only for delivering immersive audio experiences in mixed reality applications~\cite{Zaunschirm2018, Rana2019} but also for enhancing downstream tasks such as event localization~\cite{perotin2018iv}, especially when generalizing to multi-channel formats beyond binaural audio.

\begin{figure*}
    \centering
    \includegraphics[width=0.7\linewidth, trim={0 5cm 8cm 2.5cm},clip]{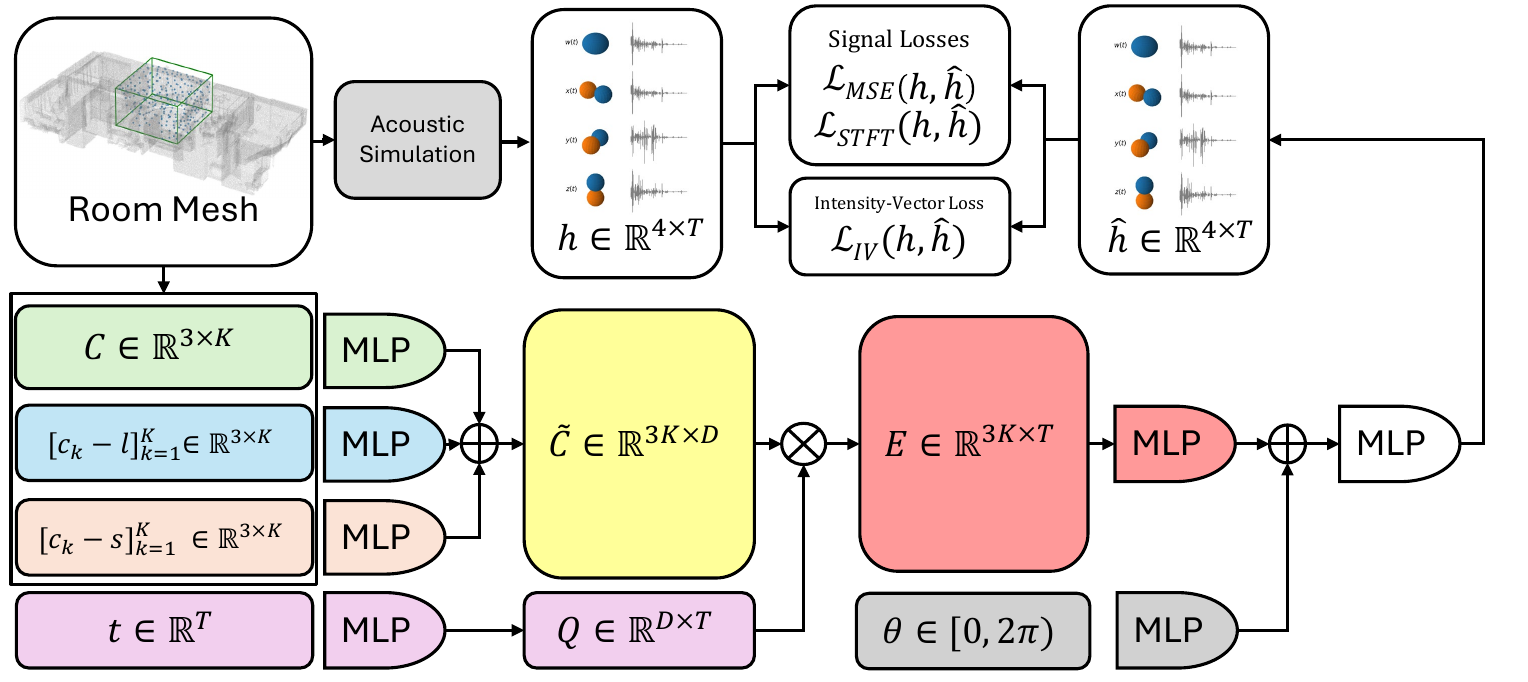}
    \caption{Overview of the proposed DANF framework. The input is the environmental context in the form of $K$ bounce points $c_k \in \mathbb{R}^3$, as well as the source and listener locations $s, l \in \mathbb{R}^3$, all of which are concatenated into spatial feature $\tilde{C}$ (denoted by operator $\oplus$) used to create a spatial-temporal encoding $E \in \mathbb{R}^{3K \times T}$. For the impulse generation module, the spatio-temporal embedding $E$ and the orientation of the listener $\theta \in [0,2\pi)$ are decoded into an Ambisonics impulse response $h \in \mathbb{R}^{4 \times T}$.}
    \label{fig:flowchart}
\end{figure*}

In this work, we propose a NF that represents spatial RIRs, allowing for direction-aware modeling of the room acoustics over continuous source and listener locations.
The proposed NF, dubbed direction-aware neural field (DANF), incorporates Ambisonic RIRs and captures the directivity of sound as illustrated in Fig.~\ref{fig:flowchart}.
Furthermore, to enhance its direction awareness, we proposed a loss function defined on an intensity vector~\cite{khaddour2013iv,perotin2018iv}.
Our experiments show that the intensity-vector loss can improve not only the sound source direction accuracy but also other single-channel RIR metrics.
In addition, we explore the adaptation of DANF to novel rooms from few measurements and demonstrate its few-shot capability.

\section{Ambisonics Room Impulse Responses}

The transfer function characterizing the physical process of sound propagation is commonly referred to as RIR. It is typically defined for given locations of the source and listener in a particular acoustic environment \cite{roginska18immersive}.
As an RIR can be treated as a linear filter, by convolving it with an anechoic sound source, we can mimic the acoustic properties of the source-listener-environment configuration (referred to as a ``scene'') in the rendered sound.
By handling RIRs, we can render the signals with any dry source in the scene.
RIRs have typically been recorded with omnidirectional microphones
giving an orientation-independent recordings for a given scene.

A popular approach to capturing a sound field at a point is to encode it into an Ambisonics format. Ambisonics captures the local pressure gradient vector by projecting it on the basis of spherical harmonics up to a given order $n$, with the increasing order corresponding to harmonics of increasing spatial resolution. The number of harmonics, and corresponding encoding channels, needed to go up to order $n$ is $(n+1)^2$, meaning the popular first-order Ambisonics encoding features 4 channels: one omnidirectional channel ($w$) and 3 figure-of-8 channels aligned with any 3 Cartesian directions (typically referred to as $x$, $y$, and $z$), as shown in top parts of Fig.~\ref{fig:flowchart}.

Each component is treated as a separate channel in a 4-channel first-order Ambisonics (FOA) signal $u(t) = [w(t), x(t), y(t), z(t)]^T$,
where $t = 1, \ldots, T$ is the time sample index, and $T$ is the number of time samples.
To gain insight on the direction-of-arrival (DoA) of sources in a scene, which tends to be associated with the azimuth and elevation of maximum intensity of the recorded signal $u \in \mathbb{R}^{4 \times T}$, an
intensity vector~\cite{khaddour2013iv,perotin2018iv} is computed from the complex time-frequency representation of each of the FOA components, for example obtained using the short-time Fourier transform (STFT) $U \in \mathbb{C}^{4 \times M \times F}$, where $M$ and $F$ denote the number of time frames and frequency bins, respectively, and $N$ is the number of channels:
\begin{equation}
    \text{IV}(U)(m,f) = \left[ \begin{array}{l}
          \text{Re}(W(m,f)^*X(m,f))\\
          \text{Re}(W(m,f)^*Y(m,f))\\
          \text{Re}(W(m,f)^*Z(m,f))
    \end{array} \right],
    \label{eq:iv}
\end{equation}
where $m$ denotes the time frame index, $f$ the frequency bin index, $^*$ the complex conjugate, and $\text{Re}$ the real component of a complex number.

\section{Methods}
\subsection{Neural Acoustic Fields}
Prior approaches in neural acoustic fields establish the problem of estimating an RIR $h \in \mathbb{R}^{N \times T}$~\cite{su22inras, liang23avnerf, liang2023nacf}.
Prior work has focused on the monaural or binaural cases, and while some have allowed for different listener orientations $\theta \in [0, 2\pi)$.
Their training methods and evaluations did not consider inter-channel directional features that would correspond to changes in orientation for a fixed source/listener. %
A neural field can then be used to estimate the binaural RIR $h \in \mathbb{R}^{2 \times T}$ as a function of the listener position $l \in \mathbb{R}^3$, the source position $s \in \mathbb{R}^3$, and some fixed environmental context $C$:
\begin{equation}
    h = \texttt{NF}(s,l,\theta,C).
\end{equation}
In many neural acoustic fields, $C$ is implicit, as a separate NF will be trained for each environment.
This environmental context can also specified via a set of ``bounce points'', as in \cite{su22inras}: each environment's geometry can be characterized by capturing a set of uniformly sampled points along the surface of each room's mesh, defining the environmental context as $C=[c_1,\dots,c_K] \in \mathbb{R}^{3 \times K}$, with each bounce point $c_k$ being normalized such that the origin is at the midpoint of the minimum/maximum coordinates as defined by the dataset metadata.

\subsection{Direction-Aware Neural Field}
We propose to go beyond binaural RIR modeling, and model an Ambisonic RIR $h \in \mathbb{R}^{4\times T}$ using a direction-aware neural field (DANF) framework:
\begin{equation}
    h=\texttt{DANF}(s,l,\theta,C),
\end{equation}
where DANF is optimized based on reconstruction losses including a direction-aware component, as described below and shown in Fig~\ref{fig:flowchart}.

We follow the architecture established in \cite{su22inras}, which can be split into a spatial embedding module and an impulse response generation module.
Using the principles of acoustic radiance transfer, the spatial embedding module learns features derived from the bounce points $c_1,\dots,c_K \in \mathbb{R}^{3}$, as well as the relative location vectors from the bounce points to the listener $[c_k-l]_{k=1}^K \in \mathbb{R}^{3 \times K}$ and from the bounce points to the source $[c_k-s]_{k=1}^K \in \mathbb{R}^{3 \times K}$.

Each of these feature sets is passed through a sinusoidal encoding and to a separate network (a 2-layer MLP for the bounce points matrix $C \in \mathbb{R}^{3 \times K}$ and a single layer MLP for the relative location vectors) for 3 separate spatial features, which are concatenated into a single spatial feature $\tilde{C} \in \mathbb{R}^{3K \times D}$.
We then choose a set of time samples $\{t_i\}_{i=1}^T$ which we encode via sinusoidal encoding and an MLP to create a set of time-domain basis functions $Q \in \mathbb{R}^{D \times T}$, which is multiplied with our concatenated spatial features to produce a spatial-temporal feature $E \in \mathbb{R}^{3K \times T}$.
This spatial-temporal feature can then be concatenated with a learned encoding of the listener orientation $\theta \in [0, 2\pi)$ at the impulse response generation module, which predicts an $N$-channel impulse response via an MLP network.

\subsection{Losses}
A common way to assess the accuracy of an estimated signal $\hat{h}$ is via direct mean squared error (MSE) loss \cite{luo22naf, su22inras}:
\begin{equation}
    \mathcal{L}_{\text{MSE}} (\hat{h}, h)= \lVert \hat{h}-h\rVert_2^2.
\end{equation}
Other typical approaches approximate human perception and utilize a time-frequency representation of the signal, such as the STFT $H$.
The magnitude STFT $|H|$ and STFT phase $\angle H$ are also often considered.
Prior approaches in NFs utilize some combination of time domain and time-frequency domain signal reconstruction losses, such as the STFT loss as defined in \cite{su22inras}, which is a combination of the magnitude loss,
$\mathcal{L}_{\text{mag}} = \lVert \lvert\hat{H}\rvert - \lvert H\rvert \rVert_1$, 
phase loss $\mathcal{L}_{\text{phase}} = \lVert \angle\hat{H} - \angle H \rVert_2 $, 
and spectral convergence $\mathcal{L}_{\text{sc}}  = \frac{\lVert\lvert\hat{H}\rvert - \lvert H\rvert \rVert_2}{\lVert \lvert H\rvert \rVert_2}$:
\begin{equation}
    \mathcal{L}_{\text{STFT}} = \mathcal{L}_{\text{mag}} + \mathcal{L}_{\text{phase}} + \mathcal{L}_{\text{sc}}.
\end{equation}

To account for interchannel information that indicates directivity, we apply a novel loss utilizing the intensity vectors of the original and reconstructed RIRs.
We integrate the intensity vector $\text{IV}(U)(m,f)$ in \eqref{eq:iv} along the frequency axis and the time axis, leading to a vector $\vec{I}(h) \in \mathbb{R}^3$.
This averaged vector corresponds the DoA a signal heard single source to a listener in an anechoic setting which should be captured in the early components of the RIR \cite{dimoulas2009improved}.
We compute the IV loss based on the cosine distance between these averaged intensity vectors for the reference and estimated RIRs:
\begin{align}
    \mathcal{L}_{\text{IV}}(\hat{h}, h) &= \frac{1}{2} \left( 1- \frac{\langle \vec{I}(\hat{h}) ; \vec{I}(h) \rangle}{\lVert\vec{I}(\hat{h})\rVert \,\lVert\vec{I}(h)\rVert}\right).
    \label{eq:ivloss}
\end{align}

\section{Experiments}

\subsection{Dataset}
We utilize Matterport3D-RGB \cite{Matterport3D} to define the acoustic environment for each of our rooms.
Because each Matterport3D building incorporates multiple connected rooms, we utilize the region metadata to slice each building into several individual rooms.
We sample 5 buildings, giving us 100 unique rooms at varying volumes.
Within each environment, we uniformly sample a grid of points at a fixed height of $1.5$ m to avoid overlapping with furniture or other objects in the room.
From this grid, for each environment, we uniformly sample 1000 source-listener pairs, with a uniformly sampled listener orientation in $[0, 2\pi)$.
We use the SoundSpaces simulation platform \cite{chen22soundspaces2} to compute RIRs in FOA for each source-listener.

\subsection{Single-Environment Performance}
To evaluate the model's performance in reconstructing accurate spatial cues for a fixed acoustic environment, we used 10 separate rooms at varying volumes.
For each room, we train and evaluate a separate DANF model using a set of 1000 source-listener pairs, with 800 randomly assigned for training and 200 for testing.

Each model was trained on a weighted sum of the MSE signal loss, STFT loss, and the intensity vector loss:
\begin{equation}
    \mathcal{L}_{\text{train}} = \mathcal{L}_{\text{MSE}} + \mathcal{L}_{\text{STFT}} + \lambda \mathcal{L}_{\text{IV}},
\end{equation}
where $\lambda$ is either zero or logarithmically spaced so that $\lambda \in \{10^\gamma \mid \gamma=-4,\dots,4\}$, to explore the effect of the intensity vector loss on the model's ability to accurately reconstruct the directivity of a given scene's RIR.

\begin{figure}
    \centering
    \includegraphics[width=0.84\linewidth]{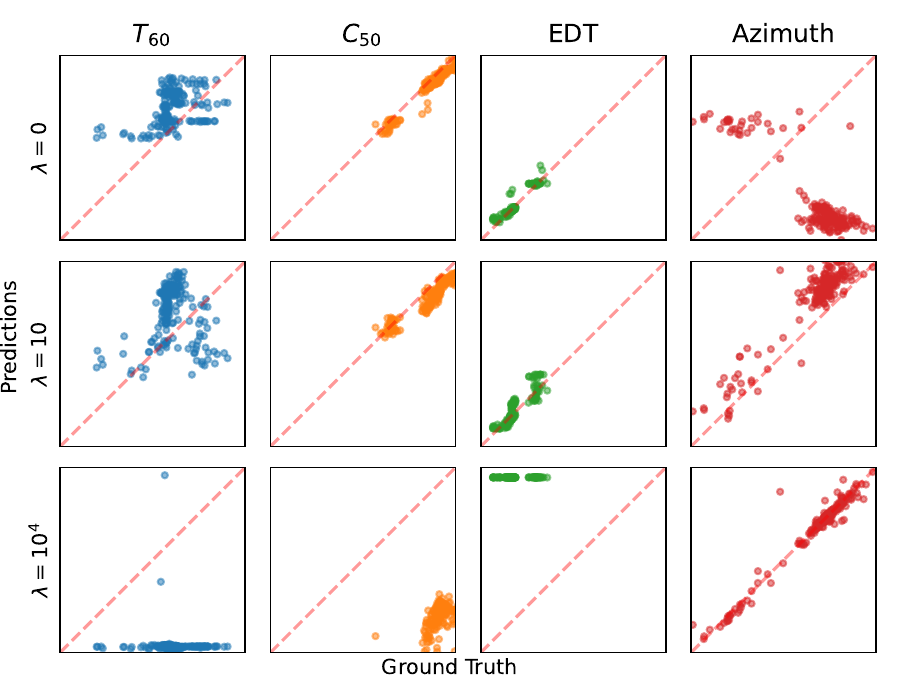}
    \caption{Single-room performance for different values of $\lambda$ (row) and metrics (columns). Each metric is normalized to $[0,1]$ for ease of visualization.}
    \label{fig:scatter}
    \vspace{-4pt}
\end{figure}

\subsection{Adaptation}
Prior work \cite{su22inras} has shown NFs' ability to be trained on multiple scenes and still provide reasonable performance for novel scenes by incorporating with room geometry.
Following the adaptation work done in \cite{chen2024RAF}, we demonstrate DANF's ability to adapt to new and unseen environments by use of fine-tuning.
We initially pre-train a model over a wide range of environments $C_{\text{pretrain}}$.
We then freeze all trainable parameters of the model and establish a new environment $C' \notin C_{\text{pretrain}}$ with a corresponding set of training and test data.
Each environment $C'$ has 200 points in its test set.
To evaluate the model's ability to adapt with limited data, we constrained the number of training samples to $[1, 2, 8, 40, 80, 160, 400, 800]$.

We then employ one of three types of fine-tuning.
For our first approach, we initialize the model from the results of our pretraining, and train the entire model, updating all parameters (\textit{Warm-Start}).
We additionally use low-rank adaptation (LoRA($r$)) \cite{hu2022lora}, adding $r$-rank weight matrices to the weights of the MLPs after the spatial temporal encoding $E$, which are learned during the fine-tuning stage from the new training set.
For a weight matrix $W \in \mathbb{R}^{i \times j}$, we define $r$-rank adaptation matricies by matrices $B \in \mathbb{R}^{i \times r}$ and $A \in \mathbb{R}^{r \times j}$, s.t. we update our weights by $W' = W + \frac{1}{r}BA$.
This allows us to tune the model by updating a constrained number of weights.

\begin{table*}[th]
    \centering
    \caption{Comparison of few-shot learning techniques based on number of parameters $N_p$, evaluating performance when fine-tuned on 1, 80, or 800 training samples.}
    \vspace{-4pt}
         \sisetup{
        detect-weight, %
        mode=text, %
        tight-spacing=true,
        round-mode=places,
        round-precision=2,
        table-format=2.2,
        table-number-alignment=center
        }
    \resizebox{\textwidth}{!}{
    \begin{tabular}{lcSS*{2}{S[table-format=3.2]}SSS[table-format=3.2]SSSSS}
        \toprule
        & &  \multicolumn{4}{c}{$1$ training example} & \multicolumn{4}{c}{$80$  training examples} & \multicolumn{4}{c}{$800$  training examples}\\ 
        \cmidrule(lr){3-6}\cmidrule(lr){7-10}\cmidrule(lr){11-14}
         & \multicolumn{1}{c}{$\mathbf{N_p}$} & \multicolumn{1}{c}{\textbf{T60}} & \multicolumn{1}{c}{\textbf{C50}} & \multicolumn{1}{c}{\textbf{EDT}} & \multicolumn{1}{c}{\textbf{DoA}} & \multicolumn{1}{c}{\textbf{T60}} & \multicolumn{1}{c}{\textbf{C50}} & \multicolumn{1}{c}{\textbf{EDT}} & \multicolumn{1}{c}{\textbf{DoA}} & \multicolumn{1}{c}{\textbf{T60}} & \multicolumn{1}{c}{\textbf{C50}} & \multicolumn{1}{c}{\textbf{EDT}} & \multicolumn{1}{c}{\textbf{DoA}} \\
        \midrule
        \textit{Zero-Shot} & 0 &  4.78 & 14.07 & 474.24 & 111.10 &  {-} & {-} & {-} & {-} &  {-} & {-} & {-} & {-}\\
        \textit{Cold-Start} & $3.5\times 10^6$ & 22.06 & 26.79 & 945.64 & 67.87 & 22.72 & 26.80 & 944.19 & 59.99 & 0.46 & 2.89 & 10.21 & 32.86 \\
        \textit{Warm-Start} & $3.5\times 10^6$ & 2.68 & 4.68 & 29.34 & 52.56 & 1.22 & 2.73 & 18.29 & 31.94 & 0.49 & 2.39 & 9.28 & 27.08 \\
        LoRA(3) & $2.9\times 10^4$ & 2.34 & 6.36 & 39.88 & 67.32 & 1.44 & 3.75 & 21.51 & 33.68 & 1.40 & 3.29 & 20.52 & 34.12 \\
        LoRA(1)& $9.6 \times 10^3$ & 4.82 & 7.07 & 88.15 & 55.43 & 1.82 & 3.76 & 27.22 & 42.32 & 1.31 & 3.86 & 22.66 & 40.67 \\
        \bottomrule
    \end{tabular}
    }
    \label{tab:nparams}
    \vspace{-4pt}
\end{table*}

Additionally, we evaluated the model with no adaptation, to examine zero-shot performance, and with a model initialized and trained from scratch (\textit{Cold-Start}).
The \textit{Warm-Start} models had a tendency to overfit when few ($<80$) training examples were provided, so we implemented early stopping on a 200 sample holdout set for those cases. %

\section{Results}

\begin{figure}
    \centering
    \includegraphics[width=0.85\linewidth, trim={0 0 0 0.5cm},clip]{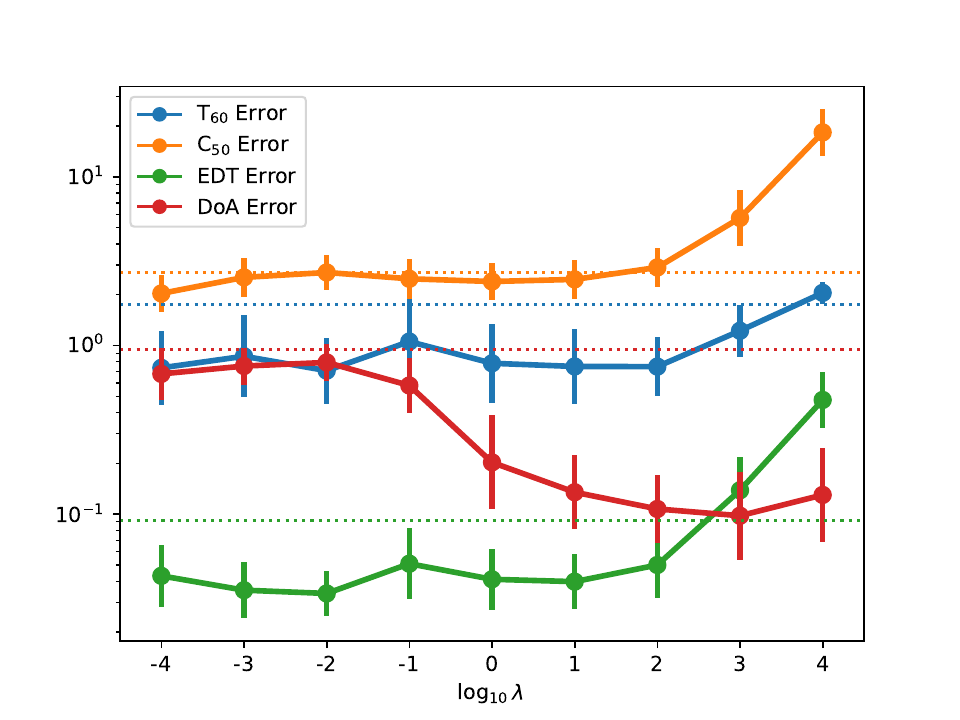}
    \caption{Model performance on single rooms with varying intensity vector loss weight $\lambda$, with the dashed line representing no intensity vector loss, or $\lambda=0$.}
    \label{fig:singleroom-plot}
    \vspace{-4pt}
\end{figure}

\subsection{Spatial Improvements}
We can see that the IV loss significantly improves DANF's ability to accurately reconstruct the directional characteristics of Ambisonic-RIRs.
Figure~\ref{fig:scatter} shows the correlations between the acoustic measures of the oracle and the rendered RIRs.
We can see the effect of adding the IV loss as the estimates of the DoA azimuth improve at increasing scales of $\lambda$.
As $\lambda$ increases, we see the model overcompensate towards accurate DoA estimates, leading the non-directional acoustic measures (T60, C50, EDT) to start to fail, most performing worse than a baseline model at $\lambda\geq 10^3$. We can see the failure of the estimates of T$_{60}$, C$_{50}$, and EDT in Fig.~\ref{fig:scatter}, as we see the model either significantly under-estimating or over-estimating each parameter, despite strong DoA azimuth performance.
On the top row, the model trained without the IV loss resulted in inaccurate estimates of the DoA azimuth.
With the IV loss with $\lambda=10$, the predicted RIRs were enforced to preserve the original IVs and showed a better correlation in the azimuth of the sound source.
The larger weight, $\lambda = 10^4$, further improved the DoA performance but deteriorated the distribution of other acoustic measures.
This result suggests that the predicted RIRs focus on the direct sound and neglect the shape of late reverberation.

Fig.~\ref{fig:singleroom-plot} depicts the acoustic measures and the DoA performance with different weights for the IV loss in \eqref{eq:ivloss}.
We can see that the addition of IV loss starts to improve model performance at $\lambda = 10^{-1}$ and continues to significantly improve DoA estimation.
The dotted lines correspond to the model without the IV loss, i.e., $\lambda = 0$.
Although too large a weight, e.g., $\lambda = 10^4$, degrades the non-directional acoustic measures (T60, C50, EDT), we can see that the intensity vector with a moderate weight, $\lambda = 10$, improves both room acoustic metrics as well as directional performance.
As we add IV loss in Table~\ref{tab:singleroom-tab}, we can see the DoA error decrease significantly, showing that without IV loss, the model either does not capture any spatial information, or incorrectly resolves the directivity of the sound field in the opposite direction.

\begin{table}[t!]
    \centering
        \caption{Per-room performance in terms of the error in T60 [\%], C50 [dB], EDT [ms] and DoA [degrees] for models trained without ($\lambda\!=\!0$) and with ($\lambda\!=\!10$) IV loss. Lower is better.}
        \vspace{-4pt}
            \label{tab:singleroom-tab}
            \sisetup{
        detect-weight, %
        mode=text, %
        tight-spacing=true,
        round-mode=places,
        round-precision=2,
        table-format=1.2,
        table-number-alignment=center
        }
        \setlength{\tabcolsep}{5pt}
        \begin{tabular}{c
            S@{\,\,}S
            S@{\,\,}S
            S[table-format=3.2]@{\,\,}S[table-format=3.2]
            S[table-format=3.0,round-precision=0]@{\,\,}S[table-format=3.0,round-precision=0]
            }
        \toprule
       & \multicolumn{2}{c}{\textbf{T60}} & \multicolumn{2}{c}{\textbf{C50}}& \multicolumn{2}{c}{\textbf{EDT}} & \multicolumn{2}{c}{\textbf{DoA}} \\
       \cmidrule(lr){2-3}\cmidrule(lr){4-5}\cmidrule(lr){6-7}\cmidrule(lr){8-9}
       $c$ & \text{$\lambda\!=\!0$} & \text{$\lambda\!=\!10$} & \text{$\lambda\!=\!0$} & \text{$\lambda\!=\!10$} & \text{$\lambda\!=\!0$} & \text{$\lambda\!=\!10$} & \text{$\lambda\!=\!0$} & \text{$\lambda\!=\!10$}  \\
       \midrule
        0 & 4.61 & 5.35 & 3.85 & 4.03 & 58.42 & 46.84 & 158 & 7\\
        1 & 1.26 & 1.10 & 2.86 & 1.74 & 16.91 & 23.55 & 68 & 35 \\
        2 & 7.42 & 0.87 & 4.34 & 3.97 & 104.94 & 38.77 & 99 & 21 \\
        3 & 0.11 & 0.08 & 0.84 & 0.90 & 23.21 & 18.91 & 57 & 35 \\
        4 & 0.24 & 0.24 & 1.80 & 4.23 & 49.14 & 81.99 & 58 & 46 \\
        5 & 0.37 & 0.39 & 3.02 & 2.29 & 10.08 & 9.75 & 157 & 19 \\
        6 & 1.09 & 1.00 & 1.95 & 2.32 & 27.29 & 27.35 & 83 & 76 \\
        7 & 0.12 & 0.10 & 1.35 & 1.09 & 31.95 & 24.57 & 112 & 41 \\
        8 & 0.19 & 0.21 & 1.11 & 0.90 & 11.46 & 13.83 & 34 & 14 \\
        9 & 0.53 & 0.60 & 4.67 & 6.57 & 234.95 & 310.06 & 85 & 69 \\
        \bottomrule
    \end{tabular}
\end{table}

\subsection{Fine-tuning}

We compare the performance of DANF pre-trained on a wide variety of rooms, and fine-tuned either by warm-starting the model on pre-trained weights, or adapting the model's pre-trained weights using LoRA.
In addition to \textit{Cold-Start}, we demonstrate the model's zero-shot performance as a baseline.
This model still leverages ``bounce points'' of the target room as its environmental context.

We can see in Table~\ref{tab:nparams} that with a large number of source-receiver paired RIRs, training a model from scratch outperforms most fine-tuning measures.
Similarly, in the low-data regime, we see that warm-starting the model and LoRA are able to more efficiently provide better estimates of a new environment's acoustic properties.
We see rank-3 LoRA achieving near equivalent performance with warm-starting the model, despite using less than $1\%$ of the number of parameters as retraining the full model, producing similar performance when fine-tuned on a single measurement (as in Table~\ref{tab:nparams}).

Furthermore, even in the low-data regime, all models do relatively well at estimating the DoA characteristics of a room; the spatial relationship between listener and source is generally unaffected by the acoustic properties of the environment they are in, so these relationships should translate relatively easily from pre-trained models.
All models adapting from pre-trained model outperform the zero-shot case, demonstrating the effectiveness of even the low-parameter LoRA approaches.

\section{Conclusion}
In this work, we presented DANF, a novel neural acoustic field model for estimating direction-aware RIRs.
To our knowledge, DANF is the first NF model estimating Ambisonic-format RIRs.
By utilizing a directional intensity vector loss, DANF is able to accurately capture not only the acoustic properties of the environment, but also the direction-dependent acoustic properties.
This is a first-of-its kind approach to capturing direction-of-arrival (DoA) with an acoustic NF.

Furthermore, we demonstrate the model's ability to adapt to new unseen environments with limited training data.
By utilizing fine-tuning strategies such as LoRA, we can see that this model's ability to adapt outperforms training-from-scratch and zero-shot approaches.

\newpage

\bibliographystyle{IEEEtran}
\bibliography{mybib}

\end{document}